\documentclass[a4paper,twoside]{article}

\usepackage{epsfig}
\usepackage{subcaption}
\usepackage{calc}
\usepackage{amssymb}
\usepackage{amstext}
\usepackage{amsmath}
\usepackage{amsthm}
\usepackage{multicol}
\usepackage{pslatex}
\usepackage{apalike}
\usepackage{algorithm2e}
\usepackage[bottom]{footmisc}
\usepackage{tikz}
\usepackage{bbm}
\usepackage{url}
\usepackage{makecell}
\usepackage{hyperref}
\usepackage{SCITEPRESS}     

\begin{document}
\title{Adversarial Malware Generation in Linux ELF Binaries via Semantic-Preserving Transformations}

\author{\authorname{Lukáš Hrdonka\sup{1}, Martin Jureček\sup{1}\orcidAuthor{0000-0002-6546-8953}}
	\affiliation{\sup{1}Faculty of Information Technology, Czech Technical University in Prague, Thákurova 9, Prague, Czech Republic}
	\email{\{hrdonluk, martin.jurecek\}@fit.cvut.cz}
}


\abstract{Malware development and detection have undergone significant changes in recent years as modern concepts, such as machine learning, have been used for both adversarial attacks and defense. Despite intensive research on Windows Portable Executable (PE) files, there is minimal work on Linux Executable and Linkable Format (ELF). In this work, we summarize the academic papers submitted in this field and develop a new adversarial malware generator for the ELF format. Using a variety of metrics, we thoroughly evaluated our generator and achieved an Evasion Rate of $67.74~\%$ while changing the confidence of the malware detector by $-0.50$ in the mean case for the dataset used. In our approach, we chose MalConv as the target classifier. Using this classifier, we found that the most successful modifications used strings typical of benign files as a data source. We conducted a variety of experiments and concluded that the target classifier appears sensitive to strings at any location within the executable file.}

\onecolumn \maketitle \normalsize \setcounter{footnote}{0} \vfill

\section{\uppercase{Introduction}}\label{sec:introduction}
IT security specialists, including academic researchers and industry professionals, have been battling against malware for decades. Over this time, the methods of malware deployment have evolved significantly~\cite{Understanding_Linux_Malware}. As a result, antimalware solutions must be continuously updated to keep pace with these advancements and effectively detect emerging threats.

Within the scope of this discussion, our primary focus will be on the generation of adversarial malware samples. In this context, attackers modify binaries to create new versions that retain the original functionality but are designed to evade detection by the target model, leading to misclassification. This technique represents a growing challenge in the cybersecurity landscape, as it can undermine the effectiveness of defense mechanisms that rely on machine learning models for threat detection. However, creating perturbations in malware binaries is a complex task because even minor modifications to the binary code can make the executable non-functional or lead to undefined behavior~\cite{Adversarial_Examples_on_Discrete_Sequences_for_Beating_Whole-Binary_Malware_Detection}.

An adversarial malware generator operates under a key restrictive condition: the program's functionality must remain unchanged. Additionally, there is the optimization to consider: the goal is to maximize the probability that the manipulated malware is misclassified as a benign file by the target detection model.

To the best of our knowledge, there is a larger number of academic papers regarding adversarial malware creation for the Windows platform (PE file format), while only a few works focus on the Linux ELF file format. However, with the massive growth of Linux usage in recent years (especially in high-performance computing, cloud services, or IoT), the research needs to investigate the possibilities of adversarial malware for the Linux ELF format. Consequently, this paper will focus mainly on the Linux ELF file format. 

The main contribution of this paper is the design of an adversarial malware generator targeting the ELF file format. Our approach is based on a genetic algorithm workflow that explores the modification space using 12 modification types and 7 data sources, increasing the diversity and effectiveness of generated samples. The work also addresses interpretability limitations of machine learning based detection through advanced logging that provides insight into the generation process. In addition, we propose evaluation metrics for adversarial generators that extend the basic evasion rate and enable a more comprehensive assessment of performance.

The rest of the paper is organized as follows. Related work done in adversarial attacks is discussed in Section~\ref{sec:relatedWork}. Section~\ref{sec:methodology} then presents the methodology and metrics used in the experiments. The proposed generator is then described in Section~\ref{sec:generatorDescription}. Section~\ref{sec:experimentalEnvironment} presents the environment used for the experiments, whose results are then described and discussed in Section~\ref{sec:experimentalResults}. At the end of the article, we outline our future work and conclude (Section~\ref{sec:conclusion}).

\section{\uppercase{Related Work}}\label{sec:relatedWork}
Despite our focus on ELF binaries, we first highlight a few works done in the field of PE files as the research dominates for these. In ~\cite{Creating_valid_adversarial_examples_of_malware}, the authors presented a black-box evasion attack using reinforcement learning
algorithms. They defined ten modification types, including appending random benign content to the end of the file or to newly created sections, removing the digital certificates information, or increasing the timestamp. They reported an Evasion Rate of 53.84~\% against the GBDT classifier, 11.41~\% against MalConv, and an average Evasion Rate of 2.31~\% against leading antivirus engines.

The work of~\cite{Malware_Makeover_Breaking_ML-based_Static_Analysis_by_Modifying_Executable_Bytes} presented a method of raw byte modification of PE files. They defined two families of transformation types -- in-place
randomization (e.g., replacing instructions with their equivalent ones of the same length, or reassigning the registers within the same functions) and code displacement (the original code is altered with the \texttt{jmp} instruction that passes control to the displaced code). They achieved evasion success rates up to 85~\% against commercial anti-viruses.

The authors of~\cite{Merlin_malware_evasion_with_reinforcement_learning} presented MERLIN -- Malware Evasion with Reinforcement LearnING. They defined 16 modification types of PE files, including adding of benign strings to the end of sections, adding import functions, or packing and unpacking the malicious file. Against MalConv, they achieved an almost 100~\% Evasion Rate, mainly because they discovered that the action type called \textit{add section strings} has a high cumulative score. They presented an Evasion Rate of 80~\% against EMBER, and 70~\% using a commercial antivirus.

In~\cite{MABMalware_A_reinforcement_learning_framework_for_blackbox_generation_of_adversarial_malware}, the authors modeled the action selection problem  using the multi-armed bandit (MAB) problem and developed MAB-Malware, which consists of two main modules -- Binary Rewriter and Action Minimizer. In Binary Rewriter, they defined eight macro-actions, including appending benign content at the end of the binary, adding random bytes to the unused space at the end of the section, or zeroing out several fields from PE files. They also implemented micro-actions, which follow some of the principles used in macro-actions but modify only a few bytes. Using Action Minimizer, they then remove unnecessary actions and replace macro-actions with micro-actions to produce a sample with only minimal changes to the binary. They presented an Evasion Rate of 74.7~\% using EMBER classifier, 97.72~\% using MalConv, and 31.99--48.3~\% using the top 3 antivirus software.

In~\cite{Learning_to_evade_static_PE_machine_learning_malware_models_via_reinforcement_learning}, OpenAI Gym code called gym-malware was released. The authors defined several manipulation types, such as adding a function to the Import Address Table that is never used, changing existing section names, or removing signer information. They also trained their own gradient boosted decision model, and presented an Evasion Rate of 24~\% against that model. Moreover, gym-malware was compared with MAB-malware in~\cite{MABMalware_A_reinforcement_learning_framework_for_blackbox_generation_of_adversarial_malware}. The work concluded that gym-malware achieved an Evasion Rate of 32.5~\% against MalConv and 15~\% against EMBER.

While moving towards ELF file format, the authors of~\cite{A_Reinforcement_Learning-Based_ELF_Adversarial_Malicious_Sample_Generation_Method} defined four major types of ELF modification methods -- add redundant data to the end of a binary file, modify the sections, add file import information, and encapsulate a file (the file content is converted into data part of \textit{Go} code). They achieved a detection rate of 25~\% for ClamAV and an Evasion Rate of 75.8~\% against VirusTotal\footnote{\url{https://www.virustotal.com/}}.

The authors of~\cite{EvilELF_Evasion_Attacks_on_Deep-Learning_Malware_Detection_over_ELF_Files} developed a tool to modify ELF binary files using deep learning algorithms. They used several types of ELF alternations, including Header Alternation, Debug Alternation, Padding Alternation, or Dynamic Extension. They reported an Evasion Rate of 76.6~\% against FireEyeNet and 8.4~\% against MalConv.

In~\cite{ADVeRL-ELF_ADVersarial_ELF_Malware_Generation_Uning_Reinforcement_Learning}, the authors developed the tool called ADVeRL-ELF to generate adversarial ELF malware using reinforcement learning. They used ResNet18, an 18-layer CNN (Convolutional Neural Network), as a target classifier. They mainly target the executable section of ELF files by adding semantic NOPs (no-operation instructions). They evaluated their tool using the IoT malware dataset (binaries compiled for the ARM architecture) and achieved a success rate of 59.5~\%. Still, they noted that their framework can be easily extended to support the x86-64 architecture as well. 

To the best of our knowledge, these are the few academic papers that discuss the offensive point of view of adversarial malware for ELF format. However, there are other works, such as~\cite{Adversarial_ELF_Malware_Detection_Method_Using_Model_Interpretation}, or~\cite{Automated_Static_Analysis_of_Linux_ELF_Malware_Framework_and_Application}, focusing primarily on the defensive aspect of adversarial malware. Furthermore, authors of~\cite{Lightweight_ELF_header_analysis_model_for_IoT_malwares_detection_based_on_machine_learning} focused on the defense against adversarial malware in lightweight ELF headers, which are used mainly in IoT.

\section{\uppercase{Background \& Evaluation Metrics}}\label{sec:methodology}
At this point, we formally define the problems of generating adversarial samples and ML-based malware detection. Our approach, tailored to ELF executable files, also benefits from the work of~\cite{louthanova2024comparison} on PE files. Further in the section, we use $\mathcal{X} \in \mathcal{D}$ to denote a malware from dataset~$\mathcal{D}$, which contains only the files classified with the malware label by the target classifier.

In our work, we developed the adversarial malware generator $\mathcal{G}$, which is defined using the formula:

\begin{equation}
	\mathcal{X}' = \mathcal{G}(\mathcal{X}) = \mathcal{X} + \delta
\end{equation}
where the adversarial perturbation $\delta$ is added to the original file $\mathcal{X}$. The output, $\mathcal{X}'$, is also ELF file that must remain unchanged in terms of execution. We then define an adversarial sample $\mathcal{X}'$ so that the malware detector classify $\mathcal{X}'$ as benign or at least reduce the confidence in malware label in comparison to $\mathcal{X}$.

To classify the ELF files, ML-based malware detector is used. For the purpose of this article, we represent this detector as a function: 
\begin{equation}\label{eq:detector}
	f( \mathcal{X} ) = (\mathcal{L}, \mathcal{P})
\end{equation} 
where $\mathcal{X}$ represents an ELF binary file, $\mathcal{L} \in \lbrace malware, benign \rbrace$ is the output label, and $\mathcal{P} \in [0.5, 1.0]$ is the probability (confidence) of that label according to detector $f$.

In our work, we use the following three metrics to evaluate the effectiveness of the proposed generator. We note that the first metric is commonly used in this field, while we developed the other two metrics to uncover additional properties of the generator. First, we use the standard Evasion Rate (ER) defined as:

\begin{equation}\label{eq:evasionRate}
	ER = \frac{\#misclassified}{total} * 100 \%
\end{equation}
where $\#misclassified$ stands for number of misclassified adversarial files by the target classifier and $total$ is a total number of files submitted to that classifier (after discarding files that were already incorrectly predicted before the actual modification).

Second, we propose an extension to the ER, the Extended Evasion Rate (EER), defined as:
\begin{equation}
	EER_{t}~=~\frac{1}{|\mathcal{D}|}~\sum_{\mathcal{X}'|\mathcal{X} \in \mathcal{D}}\mathbbm{1}_{\lbrace \alpha_{\mathcal{X}'} < t\rbrace}*~100\%
\end{equation}
for threshold parameter $t$ determining the maximal confidence included in calculation, indicator function $\mathbbm{1}$, and variable $\alpha_{\mathcal{X}'}$ defined as a confidence in malware label, thus calculated using:

\begin{equation}\label{eq:alpha}
	\alpha_{\mathcal{X}'} =
	\begin{cases}
		\mathcal{P} & f(\mathcal{X}') = ( malware, \mathcal{P}) \\
		1-\mathcal{P}  & f(\mathcal{X}') = ( benign, \mathcal{P}) 
	\end{cases}
\end{equation}
This metric was designed to better approximate the distribution of confidence in malware labels across the whole set. We propose using this metric for different values of the threshold parameter $t$ and plotting the resulting values to observe the distribution.

Third, we propose the Mean Difference in Confidence (MD) metric to observe how the confidence in malware label changes between original and adversarial malware samples in the mean case:

\begin{equation}
	MD=\frac{1}{|\mathcal{D}|}\sum_{(\mathcal{X}, \mathcal{X}')|\mathcal{X} \in \mathcal{D}}(\alpha_{\mathcal{X}'} - \mathcal{P}_{\mathcal{X}})
\end{equation}
where $\mathcal{P}_{\mathcal{X}}$ is the confidence in malware label of original malware sample $\mathcal{X}$, and $\alpha_{\mathcal{X}'}$ is computed using Equation~(\ref{eq:alpha}).

\section{\uppercase{Generator Description}}\label{sec:generatorDescription}
We describe our approach to generating malware samples in this section. We note that the generator is designed to address interpretability issues in ML algorithms by carefully logging all operations it performs. Thus, for each input file, three files are produced -- the adversarial file, \texttt{.log} file of all the operations performed (including name of the performed operation, the label of the input data, the size of the data used, and the offset to the \texttt{.data} file where the injected content resides), and the \texttt{.data} file with all the data used. 

\subsection{Modification Types}
We propose 12 modifications to ELF executables that were designed to preserve the original functionality of the binaries. These include:

\begin{itemize}
	\item Add Section. This modification adds a new section near the end of the ELF file. 
	
	\item Modify Padding Between Loadable Segments. We modify the unused space between loadable (\texttt{p\_type} is \texttt{PT\_LOAD}) segments.
	
	\item Extend Padding Between Loadable Segments. We extend the space between loadable segments to satisfy the alignment constraints (for loadable segments, page-size alignment is required~\cite{Linux_ELF_Documentation}). We use the first suitable non-zero offset for this modification.
	
	\item Rotate Loadable Segments. We move the first loadable segment after the last loadable segment. Second and forthcoming loadable segments are moved towards the beginning of the file.
	
	\item Append Benign to Malware. We append benign content to the end of the malware executable without altering its structure.
	
	\item Append Malware to Benign. We use a benign file and append block of binary zeros to meet the alignment constraints of the malware file, which is pasted afterward. The relevant fields in the ELF header are then redirected to the malware part. 
	
	\item Modify Content of Section. We modify the content of sections that have no impact on the program execution flow. However, these sections can be used by tools. To prevent these tools from producing corrupted outputs we optionally rename the section.
	
	\item Modify Content of \texttt{.strtab} Section. We modify the static symbols of ELF executable files (symbols in the \texttt{.symtab} section, which are used only by debuggers or other tools). We aim not only to add typical benign-file symbol names to malware files, but also to remove typical malware-file symbol names.
	
	\item Unregister Section. We propose removing definitions of sections that are unnecessary for the program's execution flow. Their original content is then used as a space for adversarial perturbations.
	
	\item Remove Section. We also propose removing the section entirely from the ELF file. This modification is tailored only to a selected set of sections that have no impact on the program's execution, and are not part of any loadable segments (to avoid issues with program loading).
	
	\item Change Registers. In this modification, we interchange the registers with equivalent purposes within the executable sections of the binary file. 
	
	\item Change Instructions. We propose  interchanging the instructions with the same logical (or mathematical) properties while ensuring the hardware dependencies also remain unchanged. 
\end{itemize}

\subsection{Data Sources}
For the mentioned modification types, the critical question is which data are suitable for these. We use the following data sources:

\begin{itemize}
	\item Sequence of Binary Zeros. We propose using a sequence of \texttt{0x00} bytes. Although applying this sequence is an effective way to remove the original content, the presence of it is common in ELF files (compilers often use it as a padding). 
	
	\item Random. We propose including Random data in the adversarial samples. For this data source, we assume that the content of some parts of executables may differ significantly, so that the distribution of exact byte sequences may appear random. 
	
	\item Static File. Using this data source, the actual path to the file, whose content will serve as a source for adversarial perturbations, is expected. We included two options in our implementation: rodata and text. The former contains extracted manual pages of well-known Linux commands, while the latter contains their \texttt{.text} sections. 
	
	\item Extracted Strings. We extract the strings from the \texttt{.rodata} section of benign and malware files of the specified directories. We then select only the benign strings that differ more than the defined threshold and include them in malware files.
	
	\item Extracted Symbols -- Dynamic. We extract the symbols of benign and malware files from the specified directory. We then select symbols typical of benign files to include in malware files, and also symbols typical of malware files to remove them from these files. This data source is tailored to the \texttt{.strtab} section only.
	
	\item Extracted Symbols -- Static. This data source is also tailored to the \texttt{.strtab} section. However, no symbols are extracted, hence the data source expects a static list of symbols to add and remove. 
	
	\item Benign Executable File. We use this data source exclusively for the Append Malware to Benign modification. We enable using of any valid, benign executable. However, we use \texttt{/usr/bin/xxd} primarily because of its smaller size. 
	
\end{itemize}

\subsection{Generator Workflow}

\begin{algorithm*}[ht!]
	\caption{Proposed Process of Adversarial Malware Generation.}\label{alg:generatorWorkflow}
	\SetKwInOut{Parameters}{Parameters}

	\KwIn{$generator\_input$ = directory of files serving as input to the generator.}
	
	\KwOut{$generator\_output$ = directory of adversarial files.}
	
	\Parameters{$max\_iterations$ = upper limit to number of iterative modifications to be applied (default: 2). \\
	
	\hangindent=6em
	\hangafter=1
	$enabled\_modifications\_iter$ = list of enabled iterative modifications (default: Add Section, Modify Padding Between Loadable Segments, 
	Extend Padding Between Loadable Segments, Rotate Loadable Segments,
	Append Benign to Malware, Modify Content of Section, Unregister Section, 
	Remove Section).
	
	\hangindent=6em
	\hangafter=1
	$enabled\_modifications\_single$ = list of enabled one-time modifications. (default: Modify Content of \texttt{.strtab} Section, Append Malware to Benign, Change Registers, Change Instructions).
	
	\hangindent=6em
	\hangafter=1
	$data\_sources$ = list of enabled data sources (default: Sequence of Binary Zeros, Random, Static File (rodata, text), Extracted Strings, Extracted Symbols -- Dynamic, Extracted Symbols -- Static, Benign Executable File).
	
	$modification\_attempts$ = number of times the modification is applied (default: 10). \\ 
	
	$top\_samples$ = number of files proceeding to the next iteration (default: 3). \\
	
	$target\_confidence$ = required confidence in malware label of the output file (default: 0.2). \\
	
	$samples\_output\_count$ = number of adversarial files required as output (default: 1).
	}
	
	\ForEach{$input\_file \in generator\_input$}{
		\For{$iteration \in \lbrace 0, 1, ..., max\_iterations \rbrace$}{
			Copy $input\_file$ to $generator\_temp$
			
			\ForEach{$file \in generator\_temp$}{
				\eIf{$iteration < max\_iterations$}{
				 	$modification\_list = enabled\_modifications\_iter$
					}{
					$modification\_list = enabled\_modifications\_single$
					}
					
				\ForEach{$modification \in modification\_list$}{
					\ForEach{ $data\_source \in data\_sources$}{
						Apply the $modification$ using the $data\_source$ to the $file$ ($modification\_attempts$)-times.
						
						Select modified file with the lowest confidence in malware label.
						
						Store selected file to $iteration\_best$. 
						
						Remove other files.
					}
				}
			}
			Select the $top\_samples$ with the lowest confidence in malware label from the $iteration\_best$.
			
			Store them to $generator\_temp$.
			
			Remove other files from $iteration\_best$.
		
			\If{confidence in malware of adversarial file $<$ $target\_confidence$}{
				break
			}	
		}
		
		Select the $samples\_output\_count$ samples with the lowest confidence in malware label from the $generator\_temp$.
		
		Save these files to $generator\_output$.
		
		Remove other files from $generator\_temp$.
	
	}
\end{algorithm*}
Our approach to generating malware samples is summarized in Algorithm~\ref{alg:generatorWorkflow} and can be seen as a simplified genetic algorithm, to some extent (as some of its components, such as crossover, which can easily produce corrupted executables, are not included). 

Our generator supports two types of modifications: iterative (which can be used multiple times, such as Add Section or Extend Padding Between Loadable Segments) and one-time (which can be used only once, such as Change Instructions or Append Malware to Benign). Whereas the $max\_iterations$ parameter limits the number of iterative modifications, the one-time modifications are applied only in the last additional iteration.

The data sources are often used so that two random numbers are chosen from the defined range, serving as the offset and size of the data source used for the adversarial perturbation, to avoid always using the same data. To limit the consequences of that randomization, we decided to apply each modification for the given data source multiple times and select the sample with the lowest confidence in the malware label.

\section{\uppercase{Experimental Setup}}\label{sec:experimentalEnvironment}
Our implementation is written in Python. For implementation of machine learning algorithms, we use PyTorch\footnote{\url{https://docs.pytorch.org/docs/stable/index.html}} package. We also utilize ELFFile\footnote{\url{https://pypi.org/project/elffile/}} to read and parse attributes of the ELF binary file. We use Keystone-engine\footnote{\url{https://pypi.org/project/keystone-engine/}} as an assembly framework, and the Capstone\footnote{\url{https://pypi.org/project/capstone/}} engine as a disassembly framework. 

We use the Labeled-Elfs dataset\footnote{\url{https://github.com/nimrodpar/Labeled-Elfs}} containing both malware and benign samples, which are labeled not only by package name but also by target architecture, endianness, ABI, or compiler. However, as the malware files contain only files compiled for 64-bit x86-64 architecture, we omit the other architectures from benign set. We also undersampled the benign class to create a balanced dataset. The files were split into training (64~\%), validation (16~\%), and test (20~\%) sets. Specifically, 882 files (441 malware and 441 benign) form the training set, 220 (110 malware and 110 benign) form the validation set, and 160 malware files form the testing set (we do not include the benign files in the test set).

To classify ELF executable files, we utilize MalConv~\cite{Malware_Detection_By_Eating_A_Whole_EXE}, a widely used CNN-based classifier. We use the MalConv-PyTorch implementation\footnote{\url{https://github.com/Alexander-H-Liu/MalConv-Pytorch}}, which provides effective parallel execution on a GPU (Graphics Processing Unit) using CUDA (Compute Unified Device Architecture). To prepare MalConv to classify unknown files, we train it for 5 epochs. In each epoch, we use the cross-entropy loss function and the Adam (Adaptive Moment Estimation) optimizer for learning. 

At the end of the training, we achieved an accuracy of $99.54~\%$ on the validation set. For test set, $96.87~\%$ samples was successfully detected as malware. Additionally, $90.62~\%$ of the test set samples were labeled as malware with confidence higher than $0.80$, and $84.38~\%$ of the samples with confidence higher than $0.90$.

\section{\uppercase{Experimental Results}}\label{sec:experimentalResults}

\begin{figure}[]
	\centering
	\includegraphics[width=\columnwidth]{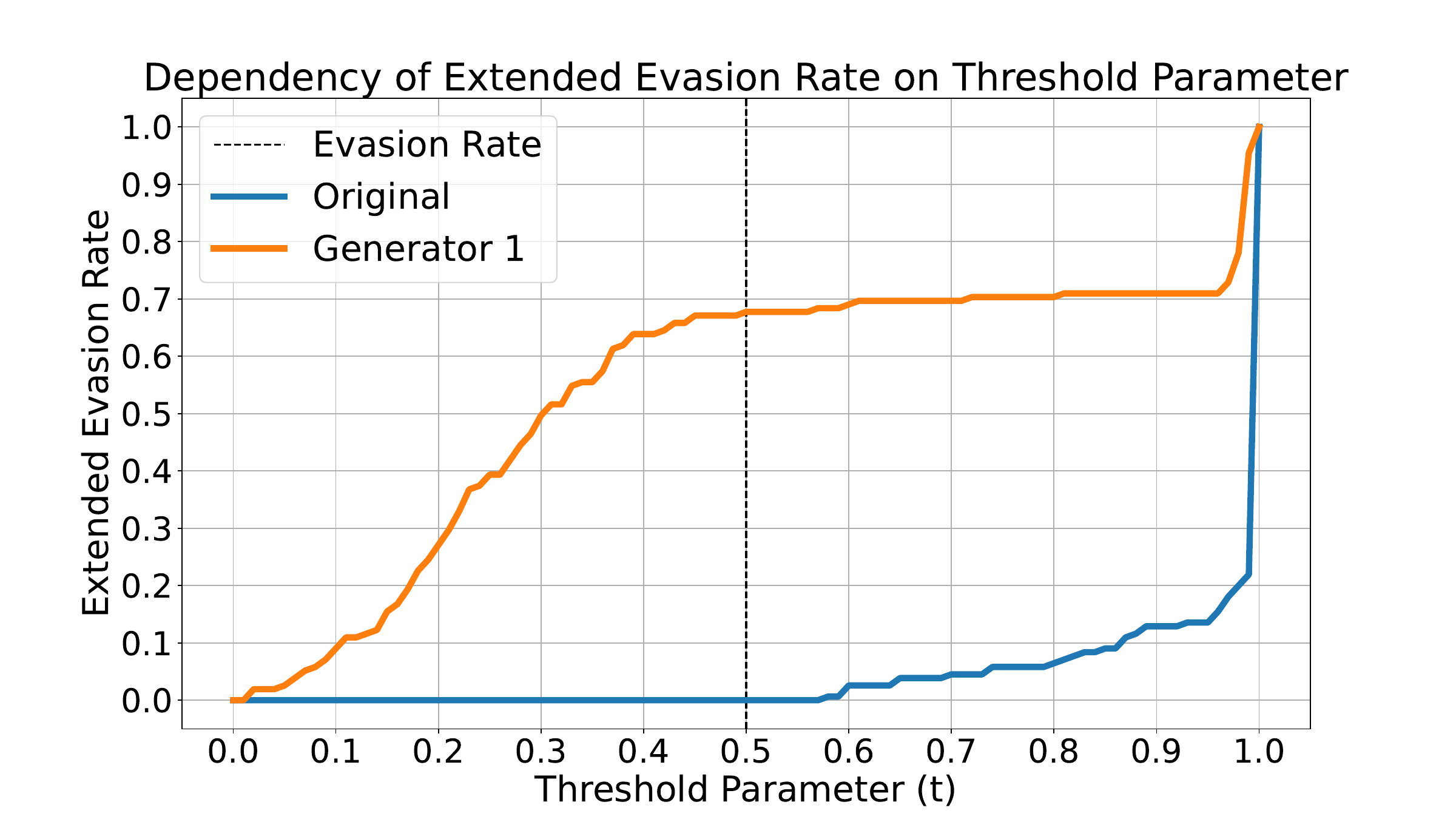}
	\caption{Experiment 1 -- Dependency of Extended Evasion Rate on Threshold Parameter. The blue line (Original) represents the malware samples before modifications are applied, whereas the orange line (Generator 1) represents the Extended Evasion Rate of the adversarial samples.}\label{img:experiment1EER}
\end{figure}

First, we evaluated the effectiveness of the proposed generator using all the modification types and data sources presented in Section~\ref{sec:generatorDescription}. More precisely, we use the default values as described in Algorithm~\ref{alg:generatorWorkflow}. Using this setup, we achieved the ER of $67.74~\%$ and changed the confidence in the malware label by $-0.5006$ in the mean case. We also calculated the EER metric for threshold values between 0.00 and 1.00 with a step size of 0.01. The output is presented in Figure~\ref{img:experiment1EER} alongside the distribution of confidence in the malware label of the original samples.

We then analyzed the \texttt{.log} files and found that only four modification types account for the majority of operations. Namely, Extend Padding Between Loadable Segments (34.61~\%), Modify Content of \texttt{.strtab} Section (28.16~\%), Add Section (16.23~\%) and Append Benign to Malware (15.99~\%), were used. We also found that Extracted Strings is used almost exclusively for modifications that do not require the more tailored data source. For the data in the \texttt{.strtab} section, Extracted Symbols -- Dynamic was used in the majority of modifications. Based on these observations, we decided to experiment with modification types and data sources in further experiments. 

In Experiment 2, we decided to limit the list of enabled modifications while keeping all the data sources enabled. We effectively divide modification types into five groups based on the operations they perform. The precise division is presented in Table~\ref{tab:experiment2Parameters} and can also be seen as a division based on the riskiness of the proposed modifications. Generator 2.1 is considered the safest because it does not alter any ELF structures. In Generators 2.2, 2.3, and 2.4, the risk of unintentionally affecting the program flow is slightly higher. The Generator 2.5 then uses modifications of executable code, which is considered the highest risk, despite our aim to optimize the modifications so that they do not affect executability.

\begin{figure}[]
	\centering
	\includegraphics[width=\columnwidth]{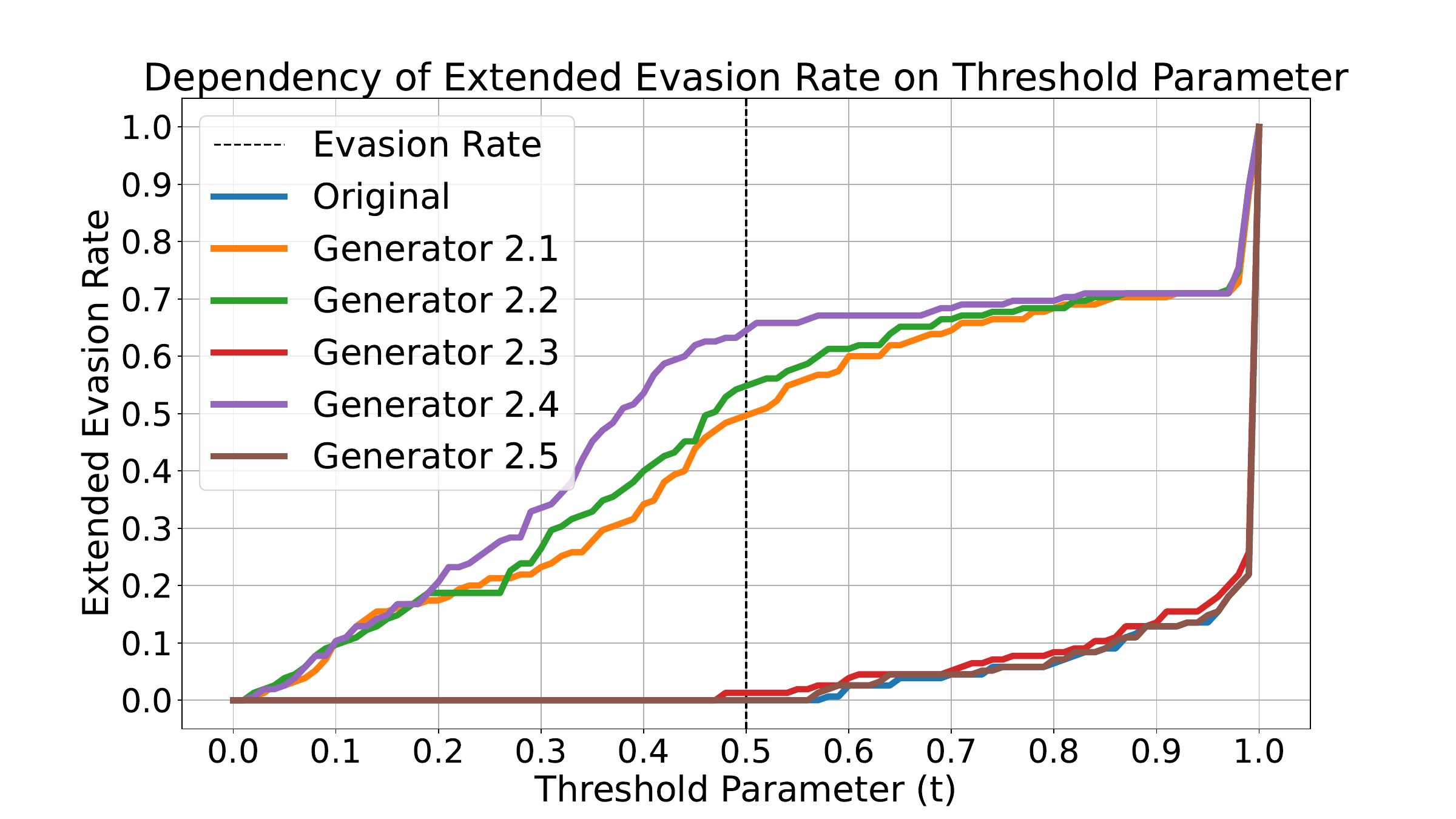}
	\caption{Experiment 2 -- Dependency of Extended Evasion Rate on Threshold Parameter.}\label{img:experiment2EER}
\end{figure}

Using this setup, we observe the highest EER for Generator 2.4, but Generators 2.2 and 2.1 provide outputs comparable to the best-performing generator. On the other hand, Generators 2.3 and 2.5 do not achieve any substantive success, as shown in Figure~\ref{img:experiment2EER}.

\begin{table*}[]
	\centering
	\caption{Experiment 2 -- Enabled Modification Types in Different Generators.}\label{tab:experiment2Parameters}%
	\begin{tabular}{@{}|l|l|@{}}
		\hline
		\textbf{Generator} & \textbf{Enabled modification types} \\  \hline
		
		Generator 2.1 & \makecell[tl]{
			Modify Padding Between Loadable Segments,
			Append Benign to Malware
		} \\  \hline
		
		Generator 2.2 & \makecell[tl]{ 
			Extend Padding Between Loadable Segments,
			Rotate Loadable Segments
		} \\  \hline
		
		Generator 2.3 & \makecell[tl]{ 
			Append Malware to Benign
		} \\  \hline
		
		Generator 2.4 & \makecell[tl]{ 
			Add Section,
			Modify Content of Section,
			Unregister Section,
			Remove Section,\\
			Modify Content of \texttt{.strtab} Section
		} \\  \hline
		
		Generator 2.5 & \makecell[tl]{ 
			Change Registers,
			Change Instructions
		} \\	 \hline
	\end{tabular}
\end{table*}

\begin{table*}[]
	\centering
	\caption{Experiment 3 -- Enabled Data Sources in Different Generators.}\label{tab:experiment3Parameters}%
	\begin{tabular}{@{}|l|l|@{}}
		\hline
		\textbf{Generator} & \textbf{Enabled data sources} \\  \hline
		
		Generator 3.1 & \makecell[tl]{
			Extracted Strings (included in the training set),
			Extracted Symbols -- Dynamic \\ (included in the training set),
			Benign Executable File
		} \\  \hline
		
		Generator 3.2 & \makecell[tl]{
			Extracted Strings (not included in the training set),
			Extracted Symbols -- Dynamic \\ (not included in the training set),
			Benign Executable File
		} \\  \hline
		
		Generator 3.3 & \makecell[tl]{ 
			Static File (rodata),
			Extracted Symbols -- Static, 
			Benign Executable File
		} \\  \hline
		
		Generator 3.4 & \makecell[tl]{ 
			Static File (text),
			Extracted Symbols -- Static,
			Benign Executable File
		} \\  \hline
		
		Generator 3.5 & \makecell[tl]{ 
			Random,
			Benign Executable File	
		} \\ \hline
	\end{tabular}
\end{table*}

\begin{figure}[]
	\centering
	\includegraphics[width=\columnwidth]{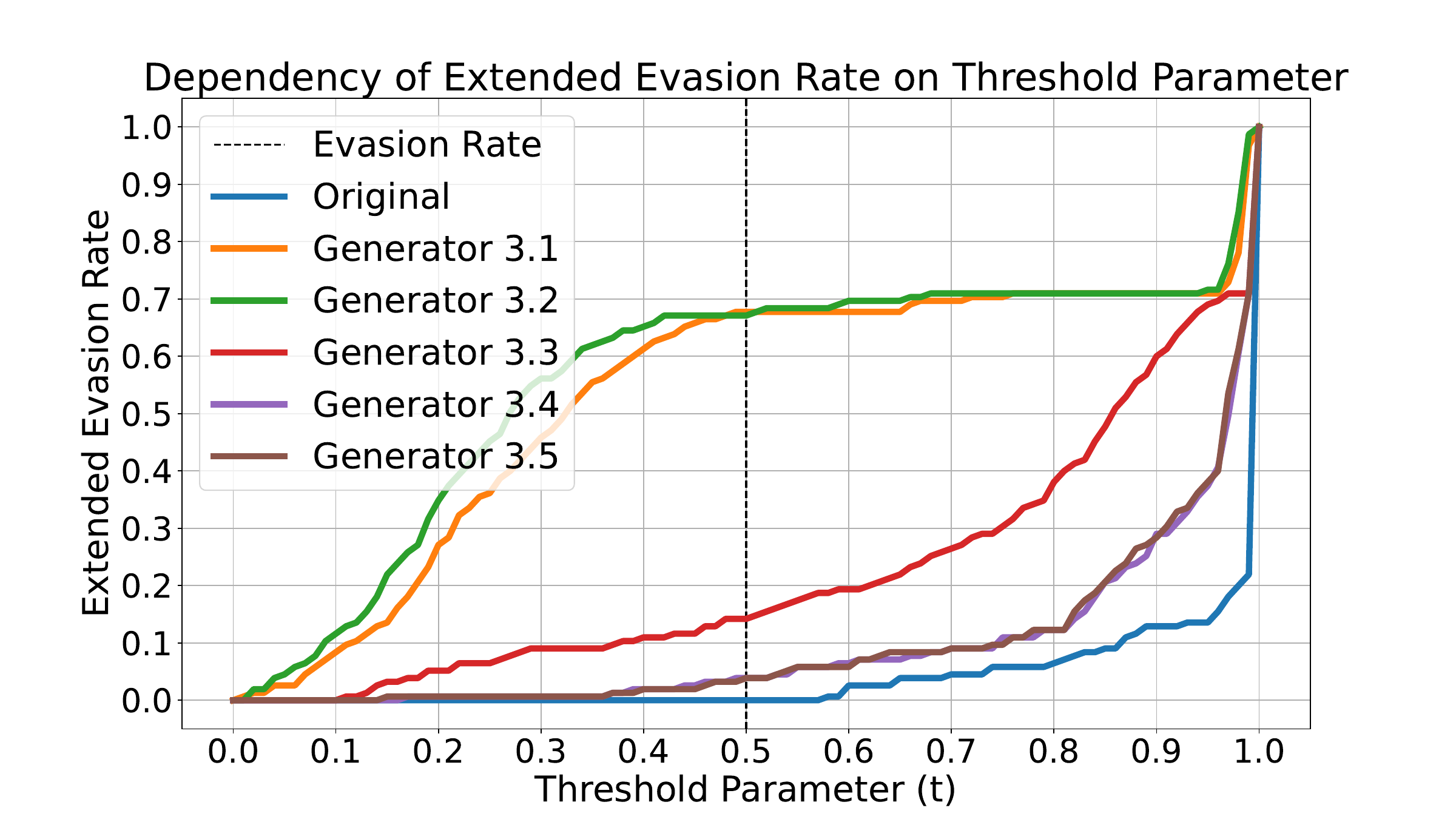}
	\caption{Experiment 3 -- Dependency of Extended Evasion Rate on Threshold Parameter.}\label{img:experiment3EER}
\end{figure}

In Experiment 3, we decided to limit the data sources used in the particular generators. In Generators 3.1 and 3.2, we enable all the data sources that extract data from malware and benign files. While Generator 3.1 extracts data from the files used in the training set, Generator 3.2 uses benign files not included in the training set and malware files intended for modification. For Generators 3.3 and 3.4, only data prepared in advance is used. The only difference between Generators 3.3 and 3.4 is the Static File data source (while Generator 3.3 uses manual pages of well-known Linux commands, Generator 3.4 uses extracted \texttt{.text} sections of the same commands). The Generator 3.5 then uses mainly the Random data source. The precise set of enabled data sources in the particular generators is presented in Table~\ref{tab:experiment3Parameters}.

Using this setup, we observe the highest EER for Generator 3.2, but Generator 3.1 remains comparable. Worse properties are then shown for Generator 3.3, which uses the contents of manual pages as primary data source. Moreover, Generators 3.4 and 3.5 achieve no substantive success as shown in Figure~\ref{img:experiment3EER}.

We also present Table~\ref{tab:experimentsComparison}, which compares the calculated Evasion Rate and Mean Difference in Confidence metrics for all the generators examined in this section. We conclude that the target classifier appears sensitive to string-based data sources but does not depend heavily on their position in the executables. However, we based our observations on the dataset that is smaller than the ones usually found in PE files.

\begin{table}[]
	\centering
	\caption{Comparison of All the Generators Presented. }\label{tab:experimentsComparison}%
	\begin{tabular}{@{}|l|c|c|@{}}
		\hline
		\textbf{Generator} & \textbf{ER} & \textbf{MD} \\  \hline
		
		Generator 1 & $67.74~\%$ &  $-0.5006$  \\ \hline
		Generator 2.1 & $49.68~\%$ & $-0.4004$ \\ \hline
		Generator 2.2 & $54.84~\%$ & $-0.4182$ \\ \hline
		Generator 2.3 & $1.29~\%$ & $-0.0092$ \\ \hline
		Generator 2.4 & $64.52~\%$ & $-0.4621$ \\ \hline
		Generator 2.5 & $0.00~\%$  & $-0.0014$ \\ \hline
		Generator 3.1 & $67.74~\%$ & $-0.4926$ \\ \hline
		Generator 3.2 & $67.10~\%$ & $-0.5216$ \\ \hline
		Generator 3.3 & $14.19~\%$ & $-0.1826$ \\ \hline
		Generator 3.4 & $3.87~\%$ & $-0.0549$ \\ \hline
		Generator 3.5 & $3.87~\%$ & $-0.0571$ \\ \hline
	\end{tabular}
\end{table}

\section{\uppercase{Conclusions \& Future Work}}\label{sec:conclusion}
This study has examined the current landscape of adversarial malware generation, with focus on ELF file format. In our work, we developed a generator of adversarial malware files for ELF file format. Our workflow is based on the genetic algorithm and accommodates 12 modification types and 7 data sources. Moreover, we aim to address interpretability issues of ML-based algorithm using advanced logging.

We conducted a variety of experiments to observe the properties of our generator. We concluded that our generator achieved the Evasion Rate of $67.74~\%$ and Mean Difference in Confidence of $-0.50$ when all the modification types and data sources were enabled. We also found that only two data sources and four modification types are used in the majority of operations. Based on that, we decided to experiment with enabling only subsets of modification types and data sources to evaluate the generator thoroughly. We divided the modification into five groups and concluded that three of the five groups produce comparable results. In a further experiment, we limited the data sources. We found that the string-based data sources were the most successful, and that the generator performed worse when other data sources were used. We concluded that the target classifier, MalConv, appears extremely sensitive to strings at any position in the executable file.

In future work, we will continue developing our generator to achieve a higher Evasion Rate within a reasonable amount of time needed for the actual generation of adversarial malware samples. We will aim to extend our generator to accommodate binaries compiled for the ARM architecture, which is primarily used in IoT. We will dive deeper into execution testing and, consequently, extract more data for dynamic analysis. We will also work on the defensive strategies to strengthen the classifiers.

\section*{\uppercase{Acknowledgements}}
This work was supported by the 2025 FIT CTU Student Summer Research Program in Prague and by the Grant Agency of the Czech Technical University in Prague, grant No. SGS26/187/OHK3/3T/18 funded by the MEYS of the Czech Republic.

\bibliographystyle{apalike}
{\small
	\bibliography{bibliography}}

\begin{thebibliography}{}

\bibitem[Anderson et~al.,
  2018]{Learning_to_evade_static_PE_machine_learning_malware_models_via_reinforcement_learning}
Anderson, H.~S., Kharkar, A., Filar, B., Evans, D., and Roth, P. (2018).
\newblock {L}earning to {E}vade {S}tatic {P}{E} {M}achine {L}earning {M}alware
  {M}odels via {R}einforcement {L}earning.

\bibitem[Cozzi et~al., 2018]{Understanding_Linux_Malware}
Cozzi, E., Graziano, M., Fratantonio, Y., and Balzarotti, D. (2018).
\newblock {U}nderstanding {L}inux {M}alware.
\newblock In {\em 2018 IEEE Symposium on Security and Privacy (SP)}, pages
  161--175.

\bibitem[Guesmi et~al.,
  2025]{Lightweight_ELF_header_analysis_model_for_IoT_malwares_detection_based_on_machine_learning}
Guesmi, H., Khalfallah, A., and Bouallegue, B. (2025).
\newblock {L}ightweight {E}{L}{F} {H}eader {A}nalysis {M}odel for {I}o{T}
  {M}alwares {D}etection {B}ased on {M}achine {L}earning.
\newblock {\em Engineering Research Express}, 7(2):025213.

\bibitem[Kosikowski et~al.,
  2023]{EvilELF_Evasion_Attacks_on_Deep-Learning_Malware_Detection_over_ELF_Files}
Kosikowski, A., Cho, D., Ninan, M., Ralescu, A., and Wang, B. (2023).
\newblock {E}vil{E}{L}{F}: {E}vasion {A}ttacks on {D}eep-{L}earning {M}alware
  {D}etection over {E}{L}{F} {F}iles.
\newblock In {\em 2023 International Conference on Machine Learning and
  Applications (ICMLA)}, pages 1702--1709.

\bibitem[Koz{\'a}k et~al.,
  2024]{Creating_valid_adversarial_examples_of_malware}
Koz{\'a}k, M., Jure{\v{c}}ek, M., Stamp, M., and Troia, F.~D. (2024).
\newblock {C}reating {V}alid {A}dversarial {E}xamples of {M}alware.
\newblock {\em Journal of Computer Virology and Hacking Techniques},
  20(4):607--621.

\bibitem[Kreuk et~al.,
  2018]{Adversarial_Examples_on_Discrete_Sequences_for_Beating_Whole-Binary_Malware_Detection}
Kreuk, F., Barak, A., Aviv-Reuven, S., Baruch, M., Pinkas, B., and Keshet, J.
  (2018).
\newblock {A}dversarial {E}xamples on {D}iscrete {S}equences for {B}eating
  {W}hole-{B}inary {M}alware {D}etection.
\newblock {\em arXiv preprint arXiv:1802.04528}, pages 490--510.

\bibitem[Louth{\'a}nov{\'a} et~al., 2024]{louthanova2024comparison}
Louth{\'a}nov{\'a}, P., Koz{\'a}k, M., Jure{\v{c}}ek, M., Stamp, M., and
  Di~Troia, F. (2024).
\newblock A {C}omparison of {A}dversarial {M}alware {G}enerators.
\newblock {\em Journal of Computer Virology and Hacking Techniques},
  20(4):623--639.

\bibitem[Lucas et~al.,
  2021]{Malware_Makeover_Breaking_ML-based_Static_Analysis_by_Modifying_Executable_Bytes}
Lucas, K., Sharif, M., Bauer, L., Reiter, M.~K., and Shintre, S. (2021).
\newblock {M}alware {M}akeover: {B}reaking {M}{L}-{B}ased {S}tatic {A}nalysis
  by {M}odifying {E}xecutable {B}ytes.
\newblock In {\em Proceedings of the 2021 ACM Asia Conference on Computer and
  Communications Security}, ASIA CCS '21, page 744–758, New York, NY, USA.
  Association for Computing Machinery.

\bibitem[Qiao et~al.,
  2023]{Adversarial_ELF_Malware_Detection_Method_Using_Model_Interpretation}
Qiao, Y., Zhang, W., Tian, Z., Yang, L.~T., Liu, Y., and Alazab, M. (2023).
\newblock {A}dversarial {E}{L}{F} {M}alware {D}etection {M}ethod {U}sing
  {M}odel {I}nterpretation.
\newblock {\em IEEE Transactions on Industrial Informatics}, 19(1):605--615.

\bibitem[Quertier et~al.,
  2022]{Merlin_malware_evasion_with_reinforcement_learning}
Quertier, T., Marais, B., Morucci, S., and Fournel, B. (2022).
\newblock {M}{E}{R}{L}{I}{N} -- {M}alware {E}vasion with {R}einforcement
  {L}earn{I}{N}g.
\newblock {\em arXiv preprint arXiv:2203.12980}.

\bibitem[Raff et~al., 2017]{Malware_Detection_By_Eating_A_Whole_EXE}
Raff, E., Barker, J., Sylvester, J., Brandon, R., Catanzaro, B., and Nicholas,
  C. (2017).
\newblock {M}alware {D}etection by {E}ating a {W}hole {E}{X}{E}.
\newblock {\em arXiv preprint arXiv:1710.09435}.

\bibitem[Ramamoorthy et~al.,
  2025]{Automated_Static_Analysis_of_Linux_ELF_Malware_Framework_and_Application}
Ramamoorthy, J., Shashidhar, N.~K., and Varol, C. (2025).
\newblock {A}utomated {S}tatic {A}nalysis of {L}inux {E}{L}{F} {M}alware:
  {F}ramework and {A}pplication.
\newblock In {\em 2025 13th International Symposium on Digital Forensics and
  Security (ISDFS)}, pages 1--5.

\bibitem[Ravi et~al.,
  2025]{ADVeRL-ELF_ADVersarial_ELF_Malware_Generation_Uning_Reinforcement_Learning}
Ravi, A., Chaturvedi, V., and Shafique, M. (2025).
\newblock {A}{D}{V}e{R}{L}-{E}{L}{F}: {A}{D}{V}ersarial {E}{L}{F} {M}alware
  {G}eneration using {R}einforcement {L}earning.
\newblock In {\em 2025 62nd ACM/IEEE Design Automation Conference (DAC)}, pages
  1--7.

\bibitem[Song et~al.,
  2022]{MABMalware_A_reinforcement_learning_framework_for_blackbox_generation_of_adversarial_malware}
Song, W., Li, X., Afroz, S., Garg, D., Kuznetsov, D., and Yin, H. (2022).
\newblock {M}{A}{B}-{M}alware: {A} {R}einforcement {L}earning {F}ramework for
  {B}lackbox {G}eneration of {A}dversarial {M}alware.
\newblock In {\em Proceedings of the 2022 ACM on Asia Conference on Computer
  and Communications Security}, ASIA CCS '22, page 990–1003, New York, NY,
  USA. Association for Computing Machinery.

\bibitem[{TIS Committee}, 2000]{Linux_ELF_Documentation}
{TIS Committee} (2000).
\newblock {T}ool {Interface} {Standard} {T}{I}{S} - {E}xecutable {a}nd
  {L}inkable {F}ormat ({E}{L}{F}) {S}pecificatoin --- linuxfoundation.org.
\newblock \url{https://refspecs.linuxfoundation.org/elf/elf.pdf}.
\newblock [Accessed 16-07-2025].

\bibitem[Xue et~al.,
  2024]{A_Reinforcement_Learning-Based_ELF_Adversarial_Malicious_Sample_Generation_Method}
Xue, M., Fu, J., Li, Z., Ni, S., Wu, H., Zhang, L.~Y., Zhang, Y., and Liu, W.
  (2024).
\newblock {A} {R}einforcement {L}earning-{B}ased {E}{L}{F} {A}dversarial
  {M}alicious {S}ample {G}eneration {M}ethod.
\newblock {\em IEEE Journal on Emerging and Selected Topics in Circuits and
  Systems}, 14(4):743--757.

\end{thebibliography}

\end{document}